\documentclass[AMA,STIX1COL]{WileyNJD-v2}

\articletype{software focus}%

\usepackage{hyperref} 

\received{26 April 2022}
\revised{6 June 2022}
\accepted{6 June 2022}

\begin{document}

\title{DTAmetasa: an R shiny application for meta-analysis of diagnostic test accuracy and sensitivity analysis of publication bias}

\author[1]{Shosuke Mizutani}

\author[2]{Yi Zhou*}

\author[1]{Yu-Shi Tian}

\author[1]{Tatsuya Takagi}

\author[1]{Tadayasu Ohkubo*}

\author[3,4]{Satoshi Hattori*}

\authormark{Mizutani \textsc{et al.}}

\address[1]{\orgname{Graduate School of Pharmaceutical Sciences}, \orgname{Osaka University}, \orgaddress{\state{Osaka}, \country{Japan}}}
\address[2]{\orgdiv{Beijing International Center for Mathematical Research}, \orgname{Peking University}, \orgaddress{\state{Beijing}, \country{China}}}
\address[3]{\orgdiv{Department of Biomedical Statistics}, \orgname{Graduate School of Medicine}, \orgname{Osaka University}, \orgaddress{\state{Osaka}, \country{Japan}}}
\address[4]{\orgdiv{Integrated Frontier Research for Open and Transdisciplinary Research Initiatives}, \orgname{Graduate School of Medicine}, \orgname{Osaka University}, \orgaddress{\state{Osaka}, \country{Japan}}}

\corres{
*Yi Zhou, Beijing International Center for Mathematical Research, Peking University, Beijing, China. \email{yzhou@pku.edu.cn}\\
*Tadayasu Ohkubo, Graduate School of Pharmaceutical Sciences, Osaka University, Osaka, Japan. \email{ohkubo@phs.osaka-u.ac.jp}\\
*Satoshi Hattori, Department of Biomedical Statistics, Graduate School of Medicine, Osaka University, Osaka, Japan. \email{hattoris@biostat.med.osaka-u.ac.jp}
}

\abstract[Abstract]{
Meta-analysis of diagnostic test accuracy (DTA) is the powerful statistical method for synthesizing and evaluating the diagnostic capacity of the medical tests and has been extensively used by clinical physicians and healthcare decision-makers. 
However, publication bias (PB) threatens the validity of meta-analysis of DTA. 
Some statistical methods have been developed to deal with PB in meta-analysis of DTA, but implementing these methods requires high-level statistical knowledge and programming skill. 
To assist non-technical users in running most routines in meta-analysis of DTA and handling with PB, we developed an interactive application, DTAmetasa.
DTAmetasa is developed with the web-based graphical user interface based on the R shiny framework. 
It allows users to upload data and conduct meta-analysis of DTA by ``point and click'' operations. 
Moreover, DTAmetasa provides the sensitivity analysis of PB and presents the graphical results to evaluate the magnitude of the PB under various publication mechanisms. 
In this study, we introduce the functionalities of DTAmetasa and use the real-world meta-analysis to show its capacity for dealing with PB. 
}


\jnlcitation{\cname{%
\author{S. Mizutani},
\author{Y. Zhou},
\author{Y-S. Tian}, 
\author{T. Takagi}, 
\author{T. Ohkubo},
and \author{S. Hattori}(\cyear{2023}),
\ctitle{DTAmetasa: an R shiny application for meta-analysis of diagnostic test accuracy and analysis of publication bias}, \cjournal{Res Syn Meth}, \cvol{2023;00:1--6}.
}}

\maketitle
\footnotetext[0]{The first two authors are contributed equally as co-first authors}

\section{Introduction}\label{sec1}

Systematic review and meta-analysis of diagnostic test accuracy (DTA) play vital roles in assessing the diagnostic accuracy of the medical test for healthcare decision-making. 
Based on the bivariate models,\cite{Rutter2001,Reitsma2005,Chu2006} 
meta-analysis of DTA synthesizes the data from multiple diagnostic studies that have the same aim and estimates the summary operating point (SOP) of sensitivity and specificity, the summary receiver operating characteristic (SROC) curve, and the area under the SROC curve (SAUC).
Since meta-analyses are always conducted with the observed data from the published studies, the presence of publication bias (PB) may be the greatest threat to the validity of the estimations.\cite{Sutton2000}

In meta-analysis of intervention studies, 
the sensitivity analysis of PB is suggested by the Preferred Reporting Items for Systematic reviews and Meta-Analyses (PRISMA).\cite{Page2021}
Meta-analysis and evaluation of PB can be implemented by accessible software such as Review Manager (RevMan, \url{https://training.cochrane.org/online-learning/core-software/revman}), 
which provides the ``point and click'' graphical user interface (GUI).
In contrast, to implement meta-analysis of DTA, it is necessary to write programming codes with statistical software such as R (\url{https://www.r-project.org/}), SAS (\url{https://www.sas.com/}), or Stata (\url{https://www.stata.com/}).
Although manuals or instructions for these software exist,\cite{Takwoingi2022} users have to spend time installing the software and learning the programming languages.
Up to date, no instruction is available to deal with PB in meta-analysis of DTA.
Some sophisticated statistical methods have been proposed recently.\cite{Hattori2018,Piao2019,Li2021,Zhou2022}
These methods adopted selection models to characterize the underlying selective publication process,
and they reported good performance in adjusting PB under certain publication mechanisms. 
However, implementing these methods requires users to have programming skills and some specific statistical backgrounds. 

Since many users can not afford the time and cost to learn how to use the software and implement the methods properly,  
there is a demand for free and accessible GUI-based tool with step-by-step guidance for implementing the meta-analysis of DTA and the analysis of PB.
Recent advances in the web-based GUIs make analytical tool development much simpler.
Based on the R Shiny framework, some applications with web-based GUIs have been developed for meta-analysis of DTA, such as
MetaDTA,\cite{Freeman2019,Patel2021} MetaBayesDTA,\cite{Freeman2019,Patel2021} Meta-DiSc,\cite{Zamora2006,Plana2022} and Meta-MUMS DTA.\cite{Sokouti2021}
These applications highly accessible via standard internet browsers and helpful for non-technical users.
{The websites of these applications are presented in Table S1 in the Supplementary Material.}
However, none of these applications supports the analysis of PB.

To break the barrier of dealing with PB in meta-analysis of DTA, we created an interactive application, DTAmetasa, based on the R shiny framework.
DTAmetasa supports the meta-analysis of DTA as well as the sensitivity analysis of PB.
Recently, Zhou et al\cite{Zhou2022} proposed the sensitivity analysis method to evaluate the impact of PB on the SROC curve or the SAUC givenbui a range of marginal selection probabilities and different selective publication mechanisms which are induced by the $t$-type statistics of the combination of sensitivity and specificity.
DTAmetasa simplifies the process of analyses and presents the results of sensitivity analysis of PB via the interactive plots.
Thus, users can visually investigate the impact of PB on the SROC curve or the SAUC and draw robust conclusions from meta-analysis of DTA.

In Section \ref{sec2}, we briefly describe the sensitivity analysis method of Zhou et al\cite{Zhou2022} and present the main functionalities of DTAmetasa. 
In Section \ref{sec3}, we use the real-world meta-analysis of DTA to illustrate how to implement DTAmetasa and interpret the results from sensitivity analysis. 
In Section \ref{sec4}, the advantages and limitations of DTAmetasa are discussed, and future extensions are presented. 

\section{Methods}\label{sec2}

To implement meta-analysis of DTA, the input data are usually the true positives (TP), true negatives (TN), false positives (FP), and false negatives (FN) of multiple primary diagnostic studies. Using such data, DTAmetasa presents the estimates of meta-analysis of DTA without accounting for PB and those accounting for PB. 

\subsection{A brief review of the likelihood-based sensitivity analysis method}

The major feature of DTAmetasa is the implementation of sensitivity analysis of PB.
In meta-analysis, it is difficult to give the explicit assumption on the mechanisms of selective publication of the observed studies; 
thus, sensitivity analysis based on the overall probability of selective publication is preferable for assessing the impact of PB.\cite{Copas1999,Copas2013}
In many published meta-analyses of DTA, the presence of PB was tested by the asymmetry of the funnel plot on the log-transformed diagnostic odds ratio (lnDOR), which suggested that the $t$-statistic of the lnDOR might be responsible for the underlying selective publication. 
Zhou et al\cite{Zhou2022} proposed the cutoff-dependent selection function to model different selective publication mechanisms and extended the likelihood-based sensitivity analysis for PB in meta-analysis of intervention studies\cite{Copas2013} into meta-analysis of DTA.
The method of Zhou et al adopted a general class of selection functions on the linear combination of logit-transformed sensitivity and specificity, which included the lnDOR as the special case.
By assigning the values of coefficients ($c_1, c_2$) of the linear combination, their method estimates of the SROC curve and the SAUC under the flexible selective publication mechanisms driven by sensitivity ($c_1=1, c_2=0$), specificity ($c_1=0, c_2=1$), lnDOR ($c_1= c_2=1/\sqrt{2}$), or the estimated combination ($\hat c_1, \hat c_2$).

The method of Zhou et al introduced the marginal selection probability, $p=P(\text{publish})$, as the sensitivity parameter, which implies the expected proportion of the published studies from the population of all studies.
Given a range of values of $p$, the SROC curves and the corresponding SAUC accounting for PB could be estimated from the conditional log-likelihood of the observed data (equation 13 in Zhou et al\cite{Zhou2022}).
Under the flexible selective publication mechanisms driven by the contrast $(c_1, c_2)$ and a range of $p$'s, 
the impact of PB could be assessed by the changes of the estimated SROC curves and SAUC.

\subsection{Interactive R shiny application: DTAmetasa}

To facilitate the implementation of the sensitivity analysis method of Zhou et al, DTAmetasa was created as the web-based GUI. 
DTAmetasa is free and accessible via \url{https://alain003.phs.osaka-u.ac.jp/mephas_web/11DTA-Meta/} by the standard internet browser software. 
The web-based GUI and statistical computations in DTAmetasa were developed using multiple R packages, {as listed in Table S2 in the Supplementary Material}. 
Figure \ref{fig1} shows the general framework and layout of DTAmetasa.
The functionalities are navigated by the tabs. 
The input of data or the choices of parameters are laid out in the left area of the panel, while the outputs are generated interactively in the right area. 
The description and the help for each panel are provided in the top area. 

The web server of DTAmetasa was built on CentOS 7 and is accessible via the MEPHAS platform (\url{https://alain003.phs.osaka-u.ac.jp/mephas/index.html}).\cite{Zhou2020}

\subsection{Functionalities and features}

DTAmetasa contains the following functionalities: (1) summary statistics of the primary diagnostic studies, (2) meta-analysis of DTA, (3) sensitivity analysis of PB, and (4) detection of PB. These functionalities are navigated by tabs and comprise a step-by-step guidance for conducting meta-analysis of DTA. 

\subsubsection{Summary of the diagnostic studies}

The ``Diagnostic Studies'' tab is used to summarize the data of the collected diagnostic studies. 
Before conducting the analysis, users need to upload the data with the columns named by TP, TN, FP, and FN.
An example was inserted in DTAmetasa to help users understand the format of data as well as the analytical procedure. 

To prepare the data, users can manually input data values in the ``Edit data'' area or choose to upload the comma-separated values (CSV) file or text document (TXT) file from the local computer. 
The uploaded data will be shown in the ``Edit data'' area and inserted in the application by clicking the update button. 
Once data are successfully uploaded, users can immediately view the original and transformed data in the right area.
As the output results, the SROC scatter plot of primary studies is presented with the choice of whether to plot the confidence intervals or the confidence regions. 
The descriptive summaries of the sensitivity or specificity, the lnDOR, and the positive or negative likelihood ratios of primary studies are presented additionally with the corresponding forest plots. 
These results are the summaries of data and cannot be reported as the estimates of meta-analysis of DTA.

\subsubsection{Meta-analysis}

The second ``Meta-analysis'' panel is used for the standard meta-analysis of DTA without accounting for publication bias. 
Two random-effects models for meta-analysis of DTA are accessible: the bivariate normal model of Reitsma et al\cite{Reitsma2005} (hereinafter, the Reitsma model) and generalized linear mixed model\cite{Chu2006,HaitaoChu2010} (hereinafter, the GLM model). 
The Reitsma model is based on the normal approximation and requires continuity correction to the zero entries.
DTAmetasa will automatically do the continuity correction by adding 0.5 to the studies with zero entries.
If data contain many zero entries, the GLM model is recommended since it does not require continuity correction.
The results of these two models are similar when the total number of subjects is large or neither sensitivity nor specificity is close to 1 or 0. 
In the Reitsma model, DTAmetasa provides the choices of different estimation methods, including the maximum likelihood (ML) estimation and the restricted maximum likelihood (REML) estimation. 
The SROC scatter plot with the SROC curves estimated by either model are output as part the results. 
The estimates of the parameters are also presented in the corresponding tab panels.

\subsubsection{Sensitivity analysis for publication bias}

To evaluate the impact of PB in meta-analysis of DTA, users can access to the ``Analysis of Publication Bias'' tab. 
According to the sensitivity analysis method of Zhou et al, users need to specify a range of values for marginal selection probabilities ($p$). 
The values of $p=1, 0.8, 0.6, 0.4$ are prespecified in DTAmetasa. 
The estimations of the SROC curves and the SAUC given different values of $p$'s will be activated by clicking the calculation buttons. 
When $p=1$, accounting for no PB, the sensitivity analysis derives the same estimates with the ML estimates of the Reitsma model in the ``Meta-analysis'' panel. 
When $p<1$, the SROC curves are estimated diversely by accounting for the different impacts of PB. 
Plots of the SROC curves and the SAUCs are presented under different selective publication mechanisms. 
When $c_1, c_2$ are estimated, the selective publication mechanism is estimated from the data of the published studies. 
When $c_1, c_2$ are assigned as equality, (1,0), or (0,1), the selective publication mechanism of PB is considered to be influenced by the significance of lnDOR, logit-transformed sensitivity, or logit-transformed specificity. 
One additional mechanism allows users to specify some otherwise mechanism. 
DTAmetasa can estimate of two types of SROC curves: the SROC curve proposed in the Reitsma model\cite{Reitsma2005} and the hierarchical SROC (HSROC) curve\cite{Rutter2001}.
The HSROC curve fixes the correlation coefficient $\rho=-1$. 
The unification of SROC and HSROC was discussed by Harbord et al\cite{Harbord2007} as well as Appendix A in Zhou et al.\cite{Zhou2022}

The results present the SROC curves and SAUC in both dynamic and static plots. 
The dynamic plots show the values of points and curves interactively and help users inspect the data and the estimates more easily,
while the static plots are formatted for saving the results. 
The estimated results are presented in the interactive tables and are available for download.
 
\subsubsection{Funnel plots for publication bias}

As the widely-used method for detecting the PB, the funnel plot method on the lnDOR is implemented additionally. 
However, users should be aware that this method may have low power in PB detection.\cite{Burkner2014,Macaskill2022}  


\section{Illustrative example}\label{sec3}


To illustrate the capacities of DTAmetasa, we reanalyzed one meta-analysis of DTA published in the Cochrane Database of Systematic Reviews (\url{https://www.cochranelibrary.com/cdsr/reviews}).
This meta-analysis aimed to assess the diagnostic test accuracy of thoracic imaging, including computed tomography (CT), chest X-ray, and ultrasound, in evaluating people with suspected COVID-19.\cite{Ebrahimzadeh2022}
We reanalyzed the result of the SROC plot of chest CT in suspected cases, as shown in Figure 5 in Ebrahimzadeh et al.\cite{Ebrahimzadeh2022}
Data of 69 studies for the SROC plot were downloaded from the paper and organized in the CSV format.

\begin{figure*}[!hbt]
\centerline{\includegraphics[width=\columnwidth]{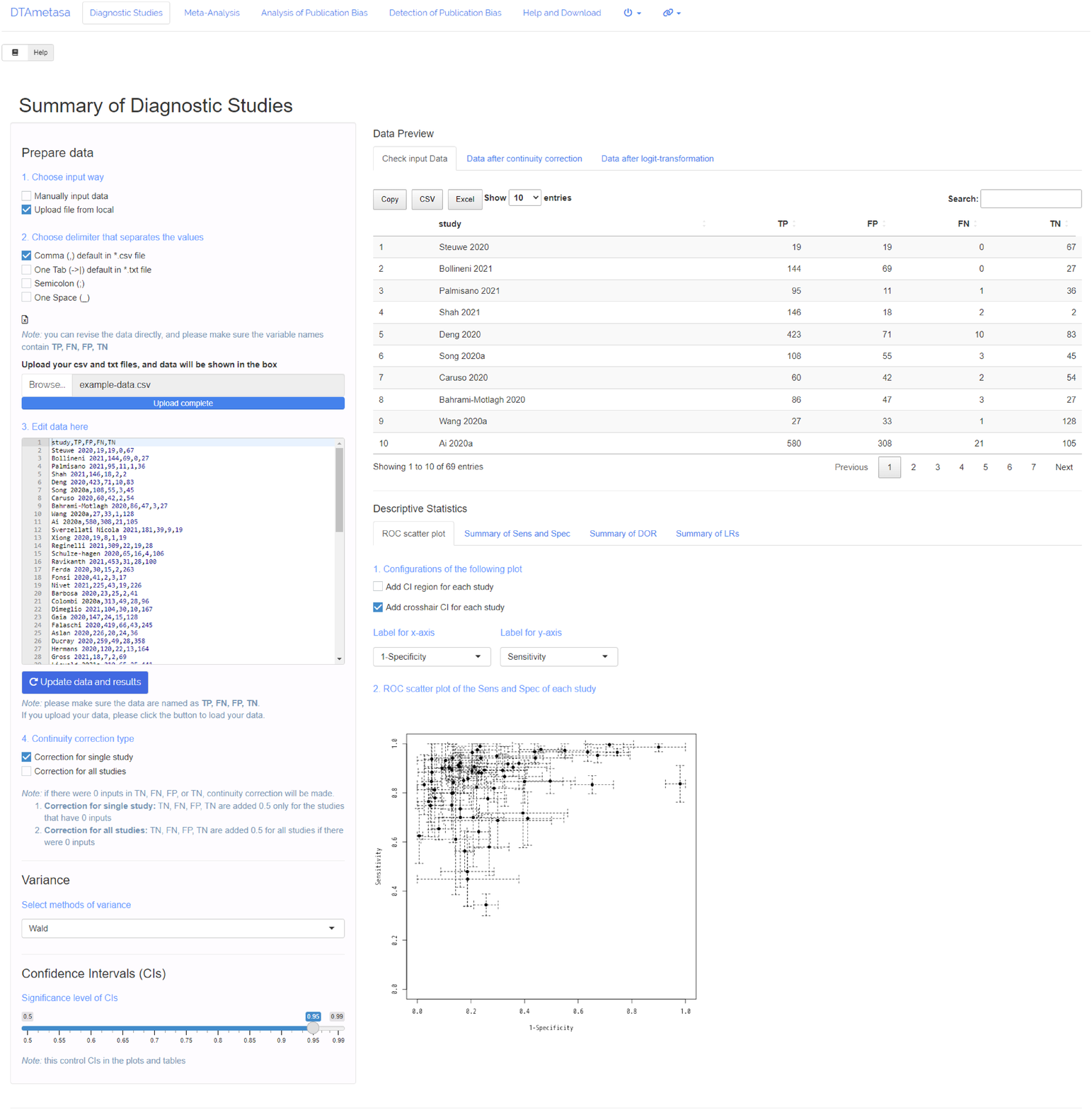}}
\caption{Screenshot: data preview and the descriptive summary of the studies in the illustrative meta-analysis.\label{fig1}}
\end{figure*}

In the first step, we chose to upload the CSV data in the ``Diagnostic studies'' panel. 
By clicking the ``Update data and results'' button, the data of 69 studies were uploaded, and the updates could be viewed in the ``Data Preview'', as shown in Figure \ref{fig1}. 
The data were automatically adjusted by the continuity correction, and the logit-transformed results were presented simultaneously. 
Under the ``Data Preview'', we could obtain the SROC scatter plot with the confidence intervals for each primary study and the other descriptive summaries.

\begin{figure*}[!hbt]
\centerline{\includegraphics[width=\columnwidth]{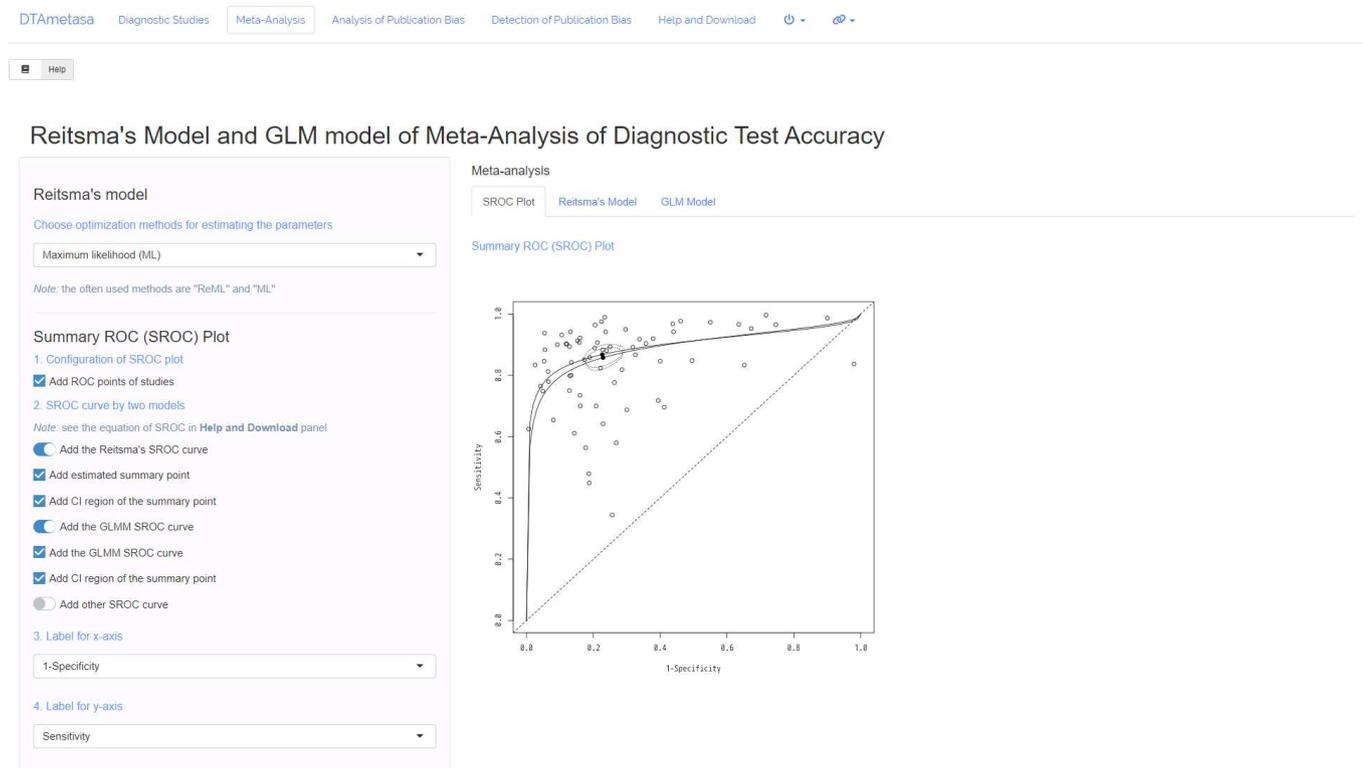}}
\caption{Screenshot: the estimated SROC curves by the Reitsma model and the GLM model without accounting for PB in the illustrative meta-analysis.\label{fig4}}
\end{figure*}

After preparing the data, we conducted the meta-analysis in the ``Meta-analysis'' tab. 
By DTAmetasa, we estimated the SROC curves by using the ML estimation in the Reitsma model and the GLM model. 
Since few studies had zero entries in the observed data, the estimated SROC curves by both models were close, as shown in Figure \ref{fig4}.
The SAUC was estimated to be 0.891 or 0.897 with the SOPs as (0.86, 0.77) or (0.87, 0.77) by the Reitsma model or the GLM model, respectively.
The estimates of the GLM model by DTAmetasa were equal to the results by MetaDTA (\url{https://crsu.shinyapps.io/dta_ma/}).\cite{Freeman2019,Patel2021}
{The screenshot of the estimations and the comparisons of the results were presented in Section 1 in the Supplementary Material.}

\begin{figure*}[!hbt]
\centerline{\includegraphics[width=\columnwidth]{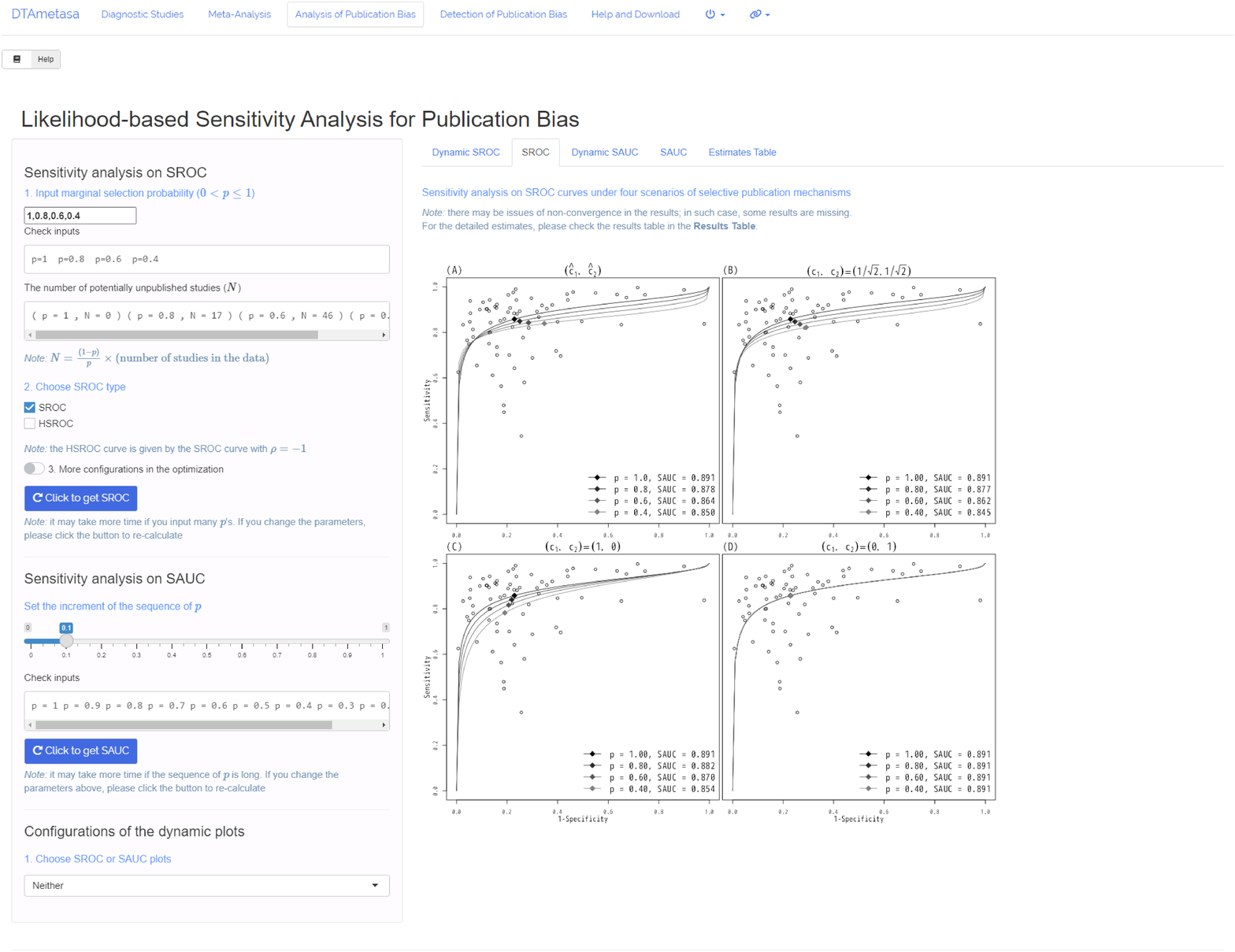}}
\caption{Screenshot: sensitivity analysis for PB on the SROC curves in the illustrative meta-analysis.\label{fig2}}
\end{figure*}

This meta-analysis did not analyze PB.\cite{Ebrahimzadeh2022}
To evaluate the impact of PB, we conducted the sensitivity analysis in the ``Analysis of Publication Bias'' tab.
We specified the marginal selection probability, $p=1, 0.8, 0.6, 0.4$, indicating that about 0, 17, 46, 103 studies were potentially unpublished based on 69 published studies.
As shown in Figure \ref{fig2}, with the increasing number of potentially unpublished studies (decreasing $p$), the SROC curves changed diversely under different scenarios of the selective publication mechanisms.
Under the selective publication mechanism estimated by the data (Figure \ref{fig2}A), 
the estimated SROC curves and SOPs changed with the estimated SAUC decreasing from 0.891 to 0.850.
The SOPs moved towards the area of low specificity and low sensitivity, indicating that studies with low sensitivity and low specificity were potentially unpublished.
Especially, when $p=0.4$, the SOPs tended to the low specificity area, which implied that with the increasing unpublished studies, the specificity might have more impact on the selective publication than the sensitivity.
Under the selective publication mechanism determined by the significance of lnDOR (Figure \ref{fig2}B), the estimated SAUC decreased to 0.845.
The trajectory of the SOPs indicated the studies with comparatively low sensitivity and low specificity were potentially unpublished.
When the selective publication mechanism was determined by the significance of sensitivity (Figure \ref{fig2}C), 
the estimated SAUC decreased to 0.854, and the trajectory of the SOPs indicated that studies with low sensitivity and high specificities were potentially unobserved.
When the selective publication mechanism was only determined by the significance of specificity (Figure \ref{fig2}D), the SROC curves did not change and no publication bias was shown.

\begin{figure*}[!hbt]
\centerline{\includegraphics[width=\columnwidth]{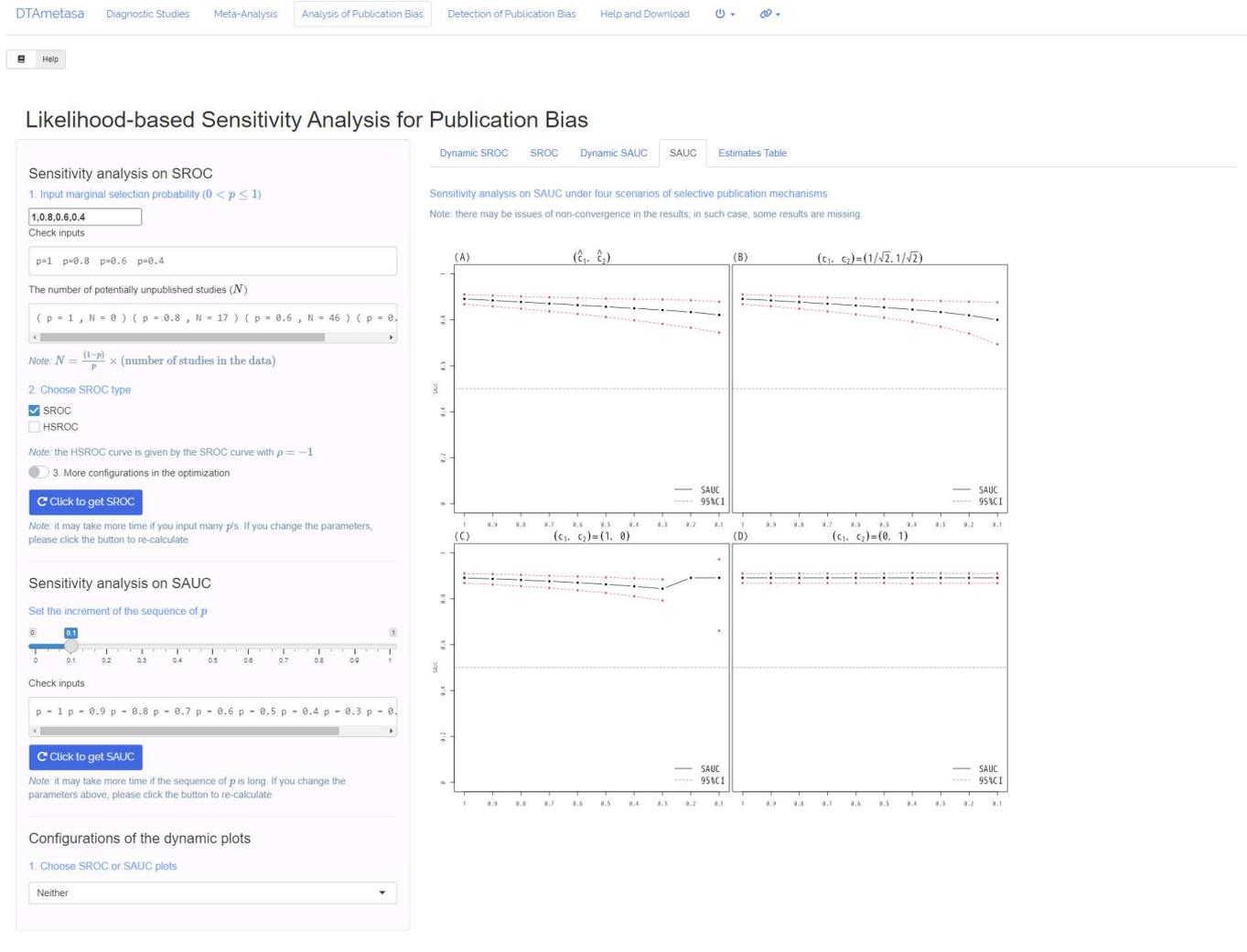}}
\caption{Screenshot: sensitivity analysis for PB on the SAUC in the illustrative meta-analysis.\label{fig3}}
\end{figure*}

The changes of the estimated SAUC were presented correspondingly in Figure \ref{fig3} given a range of $p=1, 0.9, \dots, 0.1$. 
Under all the selective publication mechanisms, the estimated SAUC did not decrease much and the confidence intervals in red were above 0.5 even when $p =0.1$. 
{The estimates of the parameters by the sensitivity analysis were presented in Section 1 in the Supplementary Material.}
The primary meta-analysis concluded the high sensitivity and specificity of chest CT for diagnosing COVID-19 in the suspected patients.\cite{Ebrahimzadeh2022}
With the sensitivity analysis of PB, we could conclude that the estimated results were robust and not sensitive to the potentially unpublished studies under various publication mechanisms.

To verify the computation results of DTAmetasa, we additionally reproduced the estimates of meta-analysis for diagnosing intravascular device-related bloodstream infection\cite{Safdar2005} used in Zhou et al.\citep{Zhou2022}
{The reproduced results were identical and were presented in Section 2 in the Supplementary Material.}


\section{Discussion}\label{sec4}

In meta-analysis of intervention studies, methods for PB have been extensively studied, and many tools have been developed for non-technical users to implement sensitivity analysis of PB. 
In contrast, no feasible software or tool is available for dealing with PB in meta-analysis of DTA. 
In meta-analysis of DTA, most users implemented the funnel plot and the trim-and-fill method to detect PB on the lnDOR by adopting the existing tools developed for meta-analysis of intervention studies.
As mentioned, the funnel plot and the trim-and-fill method may have low power\cite{Burkner2014,Macaskill2022} and can not quantify the impact of PB on the SROC curve and SAUC. 
Although sophisticated methods for adjusting PB exist,\cite{Piao2019,Li2021} these methods could not explicitly model various selective publication mechanisms. 
The sensitivity analysis method of Zhou et al\cite{Zhou2022} requires fewer parameters and presents the visual evaluation of the impact of PB under different selective publication mechanisms. 
Their method suggested that users assign different values for $c_1$ and $c_2$ as the sensitivity analysis for drawing robust conclusions.
DTAmetasa provides the GUI for implementing the sensitivity analysis method of Zhou et al and compensates for the lack of tools for sensitivity analysis of PB in meta-analysis of DTA.
DTAmetasa also helps run the routines of meta-analysis of DTA by ``point and click ''.

At present, DTAmetasa only supports the meta-analysis of DTA for the single diagnostic test without covariates. 
A natural extension would accommodate covariates and comparison of diagnostic tests. 
The sensitivity analysis method adopted in DTAmetasa was proposed based on the Reitsma model and requires continuity correction for the zero entries. 
If data are too sparse, models with no need of continuity correction or based on the bivariate binomial model should be more appropriate.\cite{Macaskill2022} 
To our best knowledge, only one method\cite{Hattori2018} for PB in meta-analysis of DTA has been proposed based on the bivariate binomial model and could be implemented in DTAmetasa in the future.

\section*{Acknowledgments}
This research was partly supported by Grant-in-Aid for Challenging Exploratory Research (16K12403) and for Scientific Research (16H06299, 18H03208) from the Ministry of Education, Science, Sports and Technology of Japan.

\subsection*{Author contributions}

MS and YZ contributed to the study conception, development of the DTAmetasa, and writing R codes.
SH and YZ contributed to the methodology.
YZ and MS contributed to drafting the manuscript. 
YST contributed to the review of DTAmetasa. 
YST, SH, TO, and TT contributed to the revision of the manuscript.
All authors approved the final version.

\subsection*{Conflict of interest}

The authors declare no potential conflict of interests.

\section*{data availability statement}
The software in this paper are available at: \url{https://alain003.phs.osaka-u.ac.jp/mephas_web/11DTA-Meta/}.
The COVID-19 data in CSV format can be accessed on GitHub: \url{https://github.com/mephas/DTAmetasa_example_data}.
The original COVID-19 data can be downloaded from \url{https://www.cochranelibrary.com/cdsr/doi/10.1002/14651858.CD013639.pub4/full}.

\section*{Supporting information}
Supplementary Material is available as part of the online article.

\section*{Highlights}
\subsection*{What is already known}

\begin{itemize}
\item With limited methods proposed to deal with publication bias (PB) in meta-analysis of diagnostic test accuracy (DTA), no feasible tool exists for implementing the methods in practice.
\item The likelihood-based sensitivity analysis method has been proposed recently for evaluating the impact of PB on the SROC curve or the SAUC by considering various mechanisms of selective publication.
\end{itemize}

\subsection*{What is new}

\begin{itemize}
\item DTAmetasa is created as the first tool for evaluating PB in meta-analysis of DTA.
\item The web-based GUI of DTAmetasa facilitates the meta-analysis of DTA as well as the sensitivity analysis of PB. 
\item DTAmetasa presents the interactive plots to help users synthesize the diagnostic capacities from multiple studies and evaluate the impact of PB in the results.
\end{itemize}

\subsection*{Potential impact for Research Synthesis Methods readers outside the authors' field}
\begin{itemize}
\item DTAmetasa provides ``point-and-click'' operations, and users can follow the step-by-step guidance to conduct meta-analysis of DTA and sensitivity analysis of PB without any installation of software.
\end{itemize}

\bibliography{refs}%

\clearpage

\clearpage

\section*{Supplementary Material of ``DTAmetasa: an R shiny application for meta-analysis of diagnostic test accuracy and sensitivity analysis of publication bias''}

\begin{table}[!htp]
\caption{R Shiny applications or software for meta-analysis of diagnostic test accuracy}
\label{table:dtaapp}
\centering
\begin{tabular}{lll}
\hline
Name      &   & Link                                              \\ \hline
\textit{Meta-DiSc 2.0}\cite{Zamora2006,Plana2022}    & Web application   & \url{https://ciberisciii.shinyapps.io/MetaDiSc2/}      \\
\textit{MetaDTA}\cite{Freeman2019,Patel2021}         &  Web application & \url{https://crsu.shinyapps.io/dta_ma/} \\
\textit{MetaBayesDTA}\cite{Freeman2019,Patel2021}         &  Web application & \url{https://crsu.shinyapps.io/MetaBayesDTA/} \\
\textit{Meta-MUMS DTA}\cite{Sokouti2021}  &   Matlab software     &     \\

\hline
\end{tabular}
\end{table}

\begin{table}[!htp]
\caption{The list of R packages used in DTAmetasa}
\label{table:package}
\centering
\begin{tabular}{ll}
\hline
Package  name          & Accessible URL                                              \\ \hline
\textit{base}           & https://CRAN.R-project.org/package=base           \\
\textit{DT}             & https://CRAN.R-project.org/package=DT             \\

\textit{readxl}         & https://CRAN.R-project.org/package=readxl      \\

\textit{ggplot2}        & https://CRAN.R-project.org/package=ggplot2        \\
\textit{graphics}       & https://CRAN.R-project.org/package=graphics       \\
\textit{grDevices}      & https://CRAN.R-project.org/package=grDevices      \\

\textit{latex2exp}      & https://CRAN.R-project.org/package=latex2exp      \\
\textit{lme4}           & https://CRAN.R-project.org/package=lme4           \\

\textit{mada}           & https://CRAN.R-project.org/package=mada           \\
\textit{magrittr}       & https://CRAN.R-project.org/package=magrittr       \\
\textit{mvmeta}         & https://CRAN.R-project.org/package=mvmeta         \\

\textit{plotly}         & https://CRAN.R-project.org/package=plotly         \\

\textit{shiny}          & https://CRAN.R-project.org/package=shiny          \\
\textit{shinythemes}    & https://CRAN.R-project.org/package=shinythemes    \\
\textit{shinyWidgets}   & https://CRAN.R-project.org/package=shinyWidgets   \\
\textit{shinyAce}       & https://CRAN.R-project.org/package=shinyAce       \\

\textit{stats}          & https://CRAN.R-project.org/package=stats          \\
\textit{utils}          & https://CRAN.R-project.org/package=utils      \\ \hline
\end{tabular}
\end{table}

\clearpage
\subsection*{Additional results of the illustrative meta-analysis}

\begin{figure*}[htp]
    \centerline{\includegraphics[width=1\columnwidth]{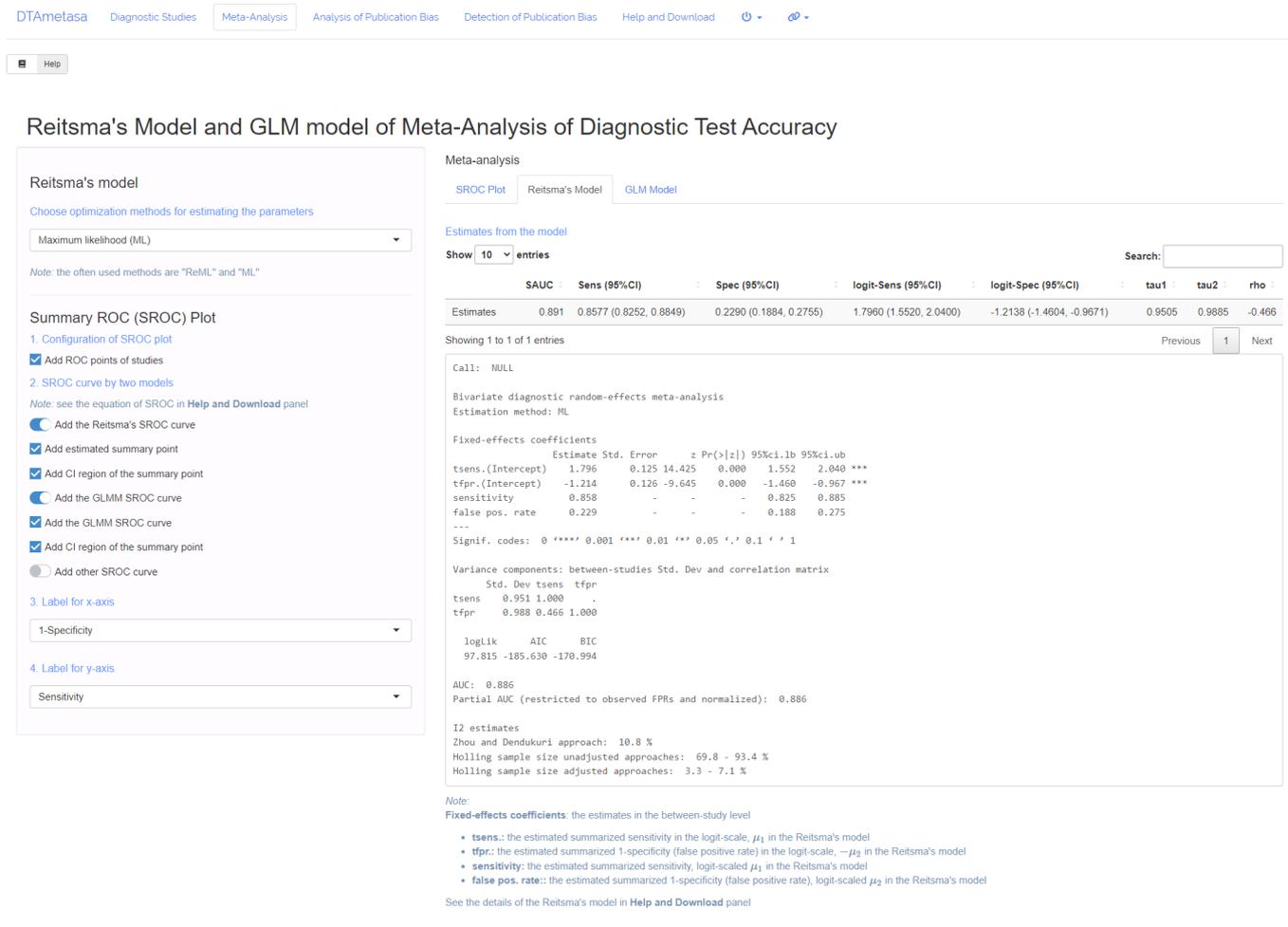}}
    \caption{Screenshot: the estimates from the Reitsma model.\label{fig2.1}}
\end{figure*}

\begin{figure*}[htp]
    \centerline{\includegraphics[width=1\columnwidth]{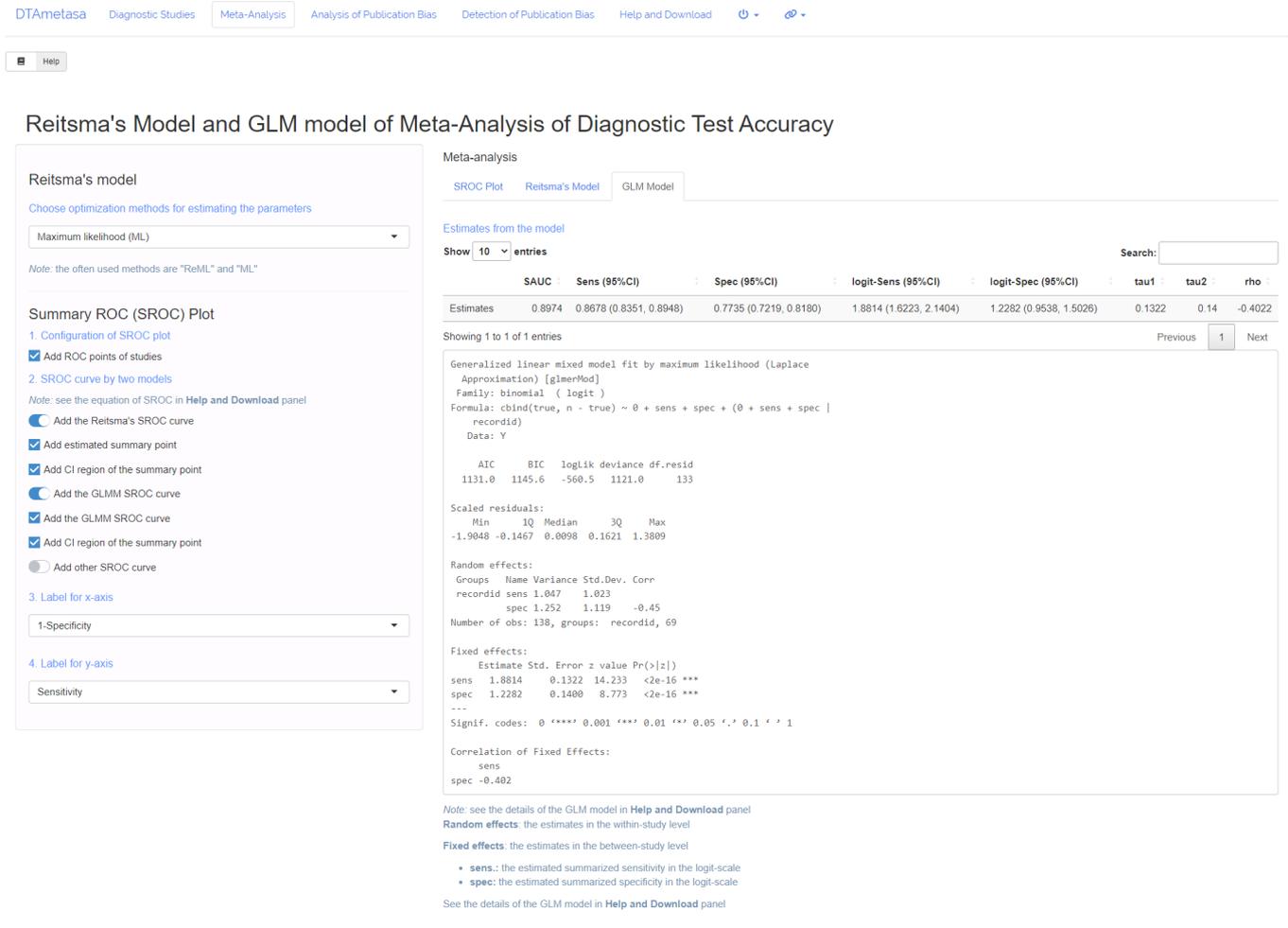}}
    \caption{Screenshot: the estimates from the GLM model.\label{fig2.2}}
\end{figure*}

\begin{figure*}[htp]
    \centerline{\includegraphics[width=1\columnwidth]{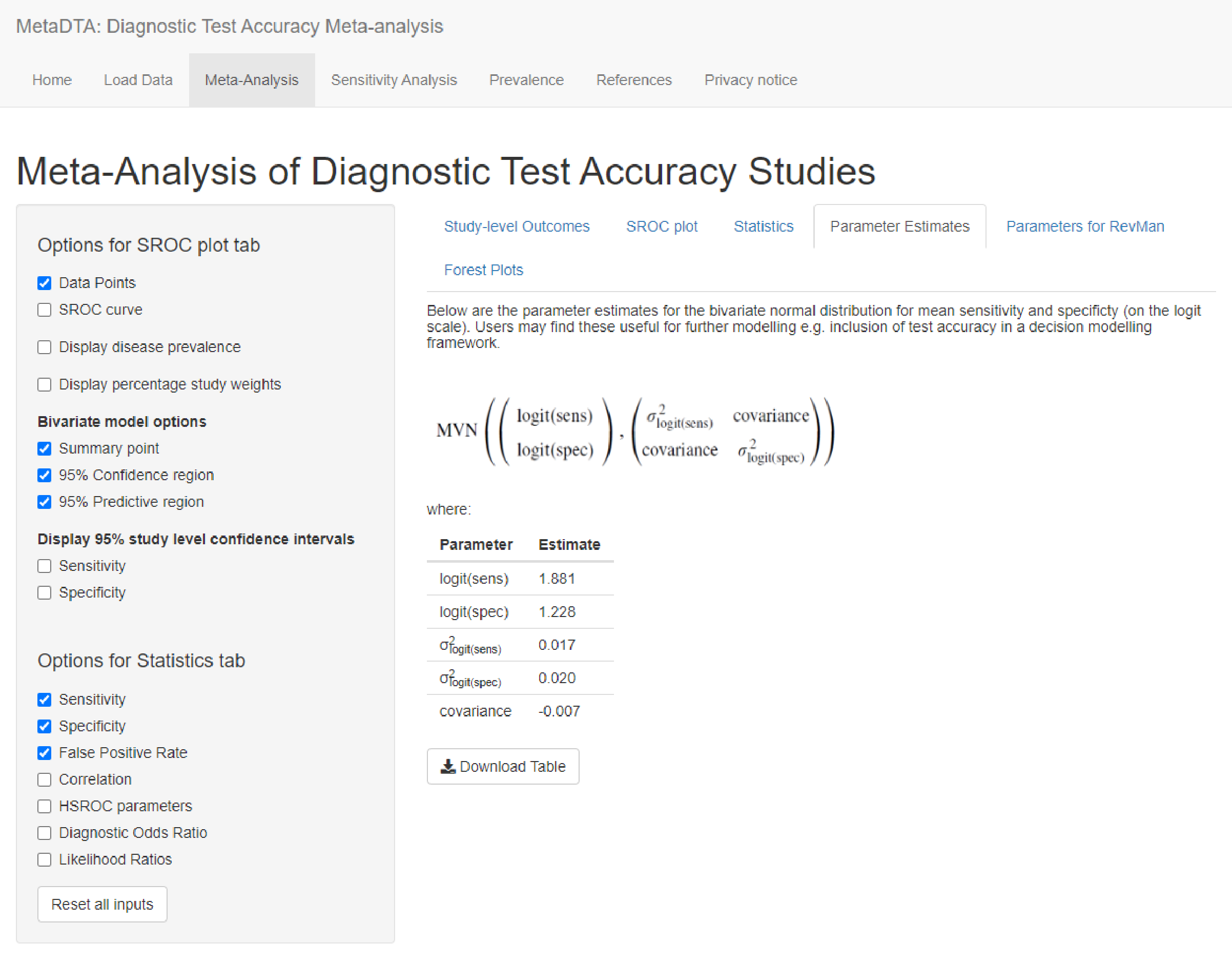}}
    \caption{Screenshot: the estimates from the GLM model in MetaDTA.\label{fig2.3}}
\end{figure*}

As mentioned in Section 3 of the main text, we presented the estimated parameters in this Section.
Figure \ref{fig2.1} shows the estimates of the Reitsma model by DTAmetasa.
Figure \ref{fig2.2} shows the estimates of the GLM model by DTAmetasa.
Figure \ref{fig2.3} shows the estimates of the GLM model by MetaDTA (\url{https://crsu.shinyapps.io/dta_ma/}).\cite{Freeman2019,Patel2021}
Results in Figure \ref{fig2.2} and Figure \ref{fig2.3} are identical.

\begin{figure*}[htp]
    \centerline{\includegraphics[width=1\columnwidth]{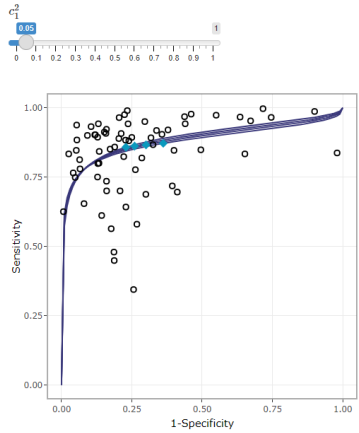}}
    \caption{Screenshot: the estimated SROC curves given $p=1, 0.8, 0.6, 0.4, 0.2$ in dynamic plots.\label{fig2.4}}
\end{figure*}

\begin{figure*}[htp]
    \centerline{\includegraphics[width=1\columnwidth]{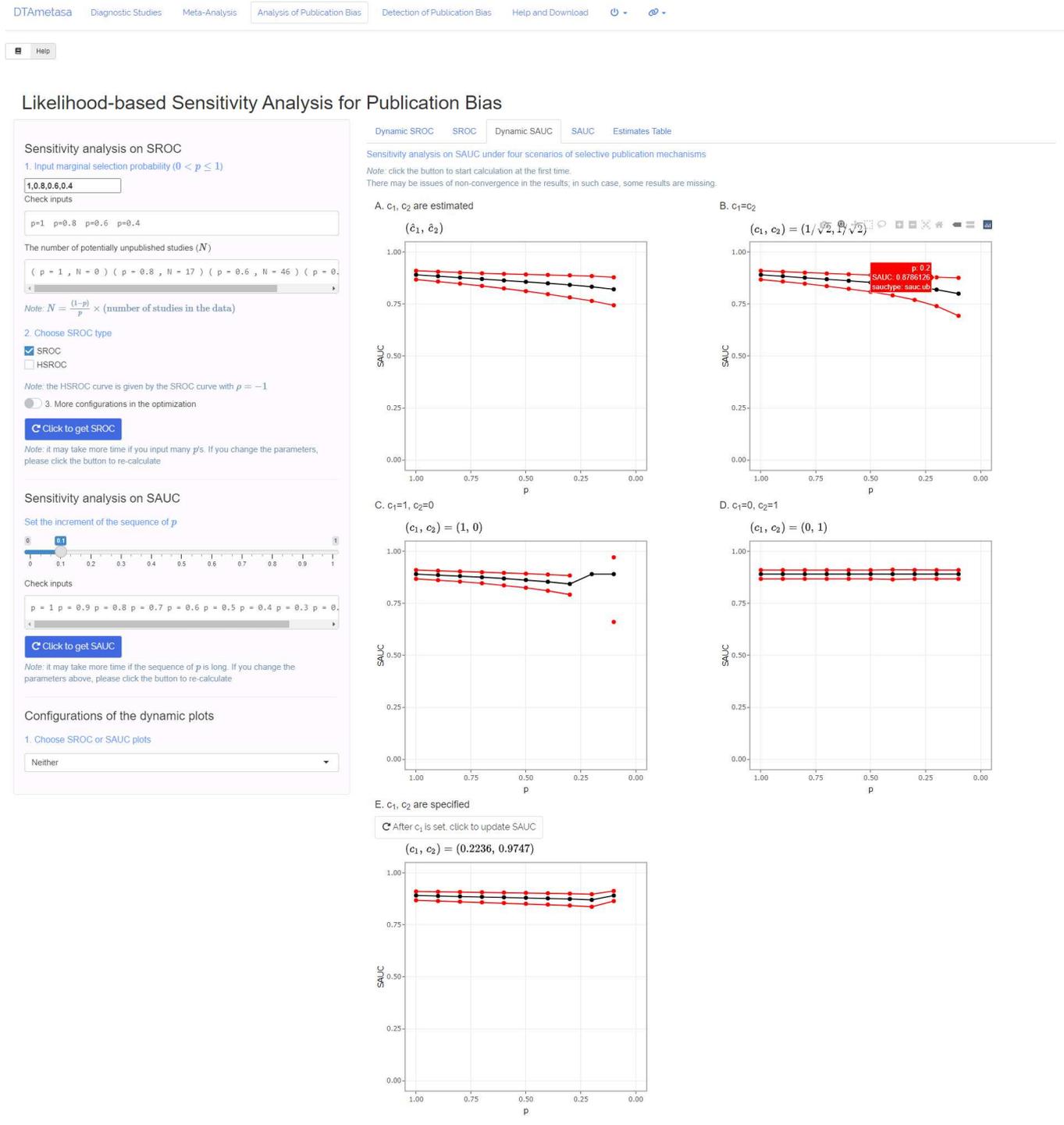}}
    \caption{Screenshot: the estimated SAUC given $p=1, 0.8, 0.6, 0.4, 0.2$ in dynamic plots.\label{fig2.5}}
\end{figure*}

\begin{figure*}[htp]
    \centerline{\includegraphics[width=1\columnwidth]{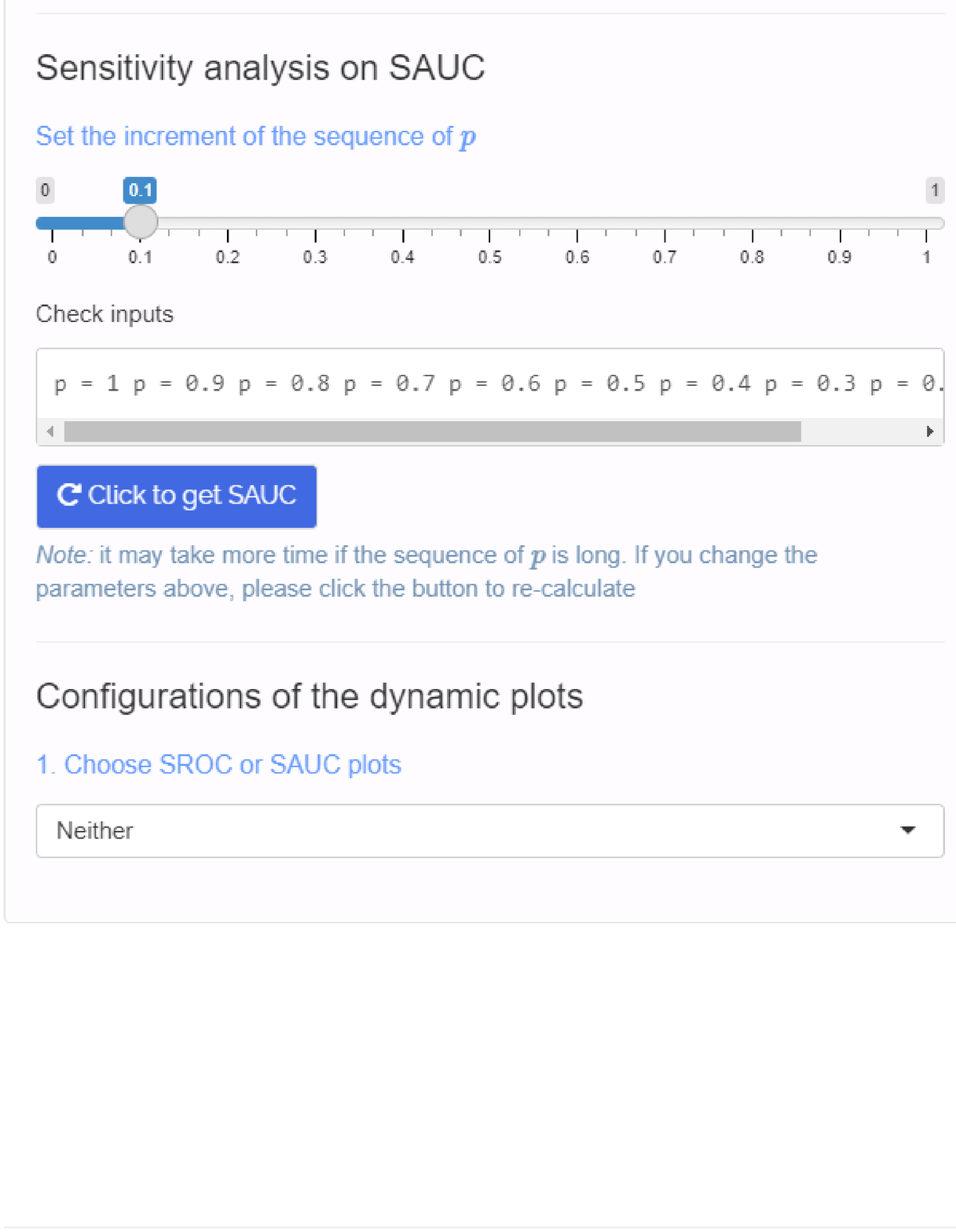}}
    \caption{Screenshot: the estimates of sensitivity analysis given $p=1, 0.8, 0.6, 0.4$.\label{fig2.6}}
\end{figure*}

Given $p=1, 0.8, 0.6, 0.4$, Figure \ref{fig2.4} gives the estimated SROC curves;
Figure \ref{fig2.5} gives the estimated SAUC;
Figure \ref{fig2.6} give the estimates of the parameters.

\subsection*{Reproducible results of meta-analysis of IVD}
As mentioned at the end of Section 3 in the main text, we presented the screenshots of reproducible results of the meta-analysis of intravascular device-related bloodstream infection\cite{Safdar2005}.
The results are identical with the estimations in Table S1 in Zhou et al.\cite{Zhou2022}

\begin{figure*}[htp]
    \centerline{\includegraphics[width=1\columnwidth]{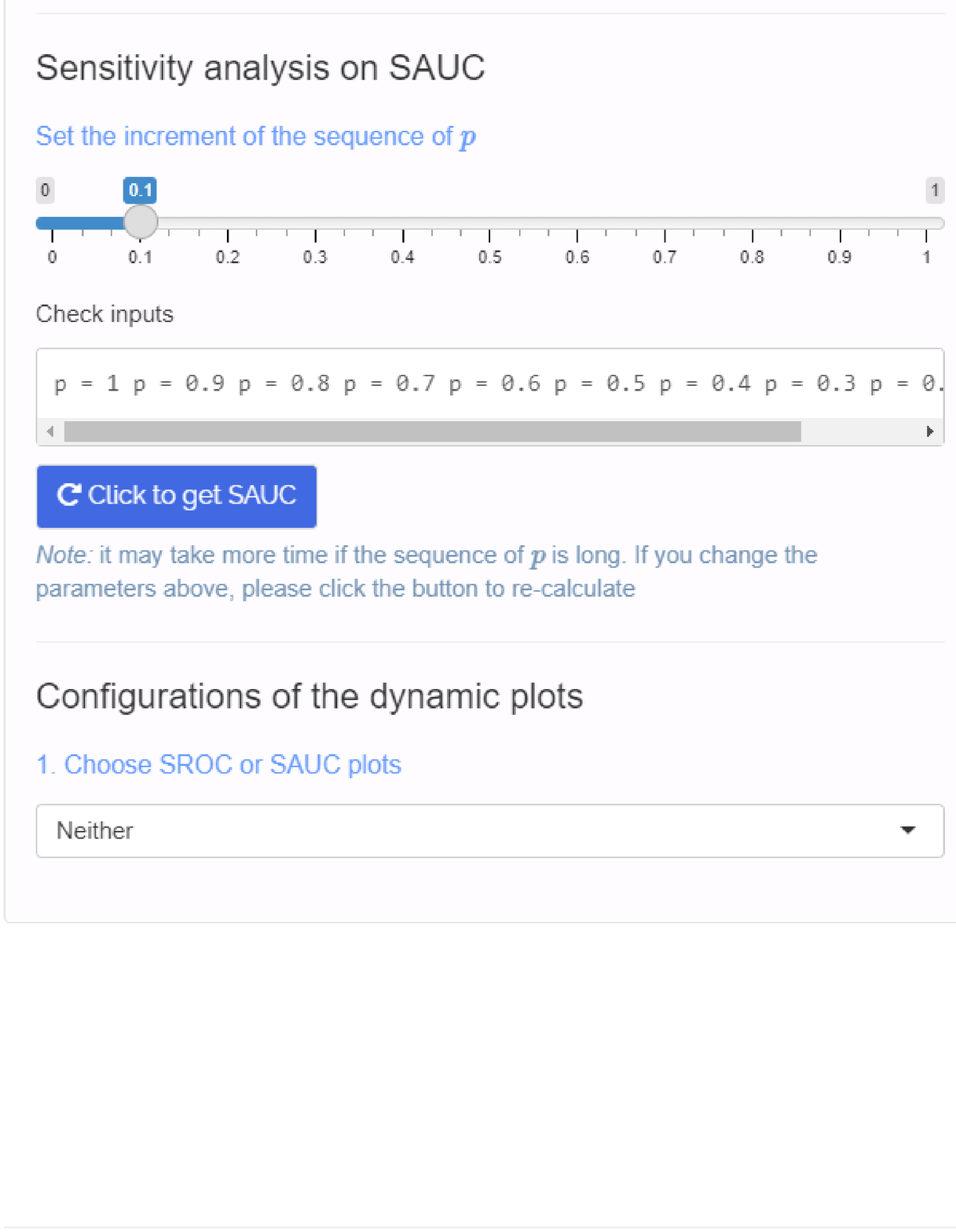}}
    \caption{Screenshot: the reproducible estimations by DTAmetasa.\label{fig3.1}}
\end{figure*}

\end{document}